\documentclass[12pt]{article}
\usepackage{graphicx}
\usepackage{psfrag}
\usepackage{epstopdf}
\usepackage{latexsym}
\usepackage{slashed}
\usepackage{amsmath}

\begin{document}

\begin{titlepage}
\null\vspace{-62pt}

\pagestyle{empty}
\begin{center}

\vspace{1.0truein} {\Large\bf Classical interactions in quantum field theory}

\vspace{1in}
{\large Dimitrios Metaxas} \\
\vskip .4in
{\it Department of Physics,\\
National Technical University of Athens,\\
Zografou Campus, GR 15773 Athens, Greece\\
 metaxas@mail.ntua.gr}\\

\vspace{.5in}
\centerline{\bf Abstract}

\baselineskip 18pt
\end{center}

I review the formalism, Feynman rules, and combinatorics that constrain a field to propagate ``classically", strictly in tree diagrams, either by itself, or interacting with other, purely quantum fields. 

The perturbation theory is reorganized by virtue of the linear terms that introduce the constraints via Lagrange multipliers, generalizing and giving results that cannot be obtained with the standard procedures which start at the quadratic terms.

I apply the formalism to a theory of an $O(N)$-symmetric quantum field interacting with a ``classical" scalar field via cubic interactions in six spacetime dimensions.
Using the renormalization group, I examine the effective potential, symmetry breaking with radiative corrections, the fixed points in $d=6-\epsilon$ dimensions, and compare with other works.

Other possible generalizations and applications of the formalism are also discussed.

\end{titlepage}
\newpage
\pagestyle{plain}
\setcounter{page}{1}
\newpage

\section{Introduction}

In a previous work \cite{dkm1} I derived the Feynman rules that are relevant when a scalar field is constrained to propagate classically (in tree diagrams) via a Lagrange multiplier, $\lambda$. The linear nature of the constraint leads to a re-organization of the perturbation theory, which is reviewed here in Sec.~2. I use the example of scalar fields with quartic interactions in four spacetime dimensions, and I discuss the methods of calculation of the effective action when linear terms are important.

In Sec.~3, I apply the formalism to a theory of an $O(N)$-symmetric quantum field, interacting with a ``classical" scalar field, via cubic interactions in six spacetime dimensions, that has been investigated in other works \cite{klebanov, souza} with respect to dimensional transmutation and the $\epsilon$-expansion, which are also examined here. The matching with a lower-dimensional conformal field theory (CFT) now is not exact beyond the leading order in the $\frac{1}{N}$-expansion, but the infrared (IR) fixed points appear for much smaller values of $N$. Symmetry breaking (for the classical field) via radiative corrections (of the quantum field) and dimensional transmutation are also present here.

In Sec.~4, I compare my formalism with other works that claim to describe classical behavior. In Sec.~5, I discuss other possible generalizations and applications, as well as possible relations to conceptual problems of measurement, and conclude with some comments.

\section{Propagation in tree diagrams}

The textbook example of a scalar field, $\phi$, in quantum field theory, with action $ S_1(\phi) = \int  d^4 x \,{\cal L} $, and 
Lagrangian ${\cal L}_1 = \frac{1}{2} (\partial_\mu \phi)^2-\frac{1}{2} m_1^2 \phi^2  -\frac{g}{4!} \phi^4 $
can be compactly described in the path integral
\begin{equation}
Z(J) = e^{i W(J)} = \int [ d\phi ] e ^{i S_1(\phi, J)},
\end{equation}
with $S_1(\phi, J)= \int d^4 x \,( {\cal L}_1 + J \phi )$. Here, $m$ and $g$ are the mass and coupling constant parameters in four  spacetime dimensions, and the metric is Lorentzian, mostly negative.
$Z[J]$ is the generating functional for all the correlation functions of the quantum theory, and $W[J]$ for the connected ones.

It was realized in \cite{dkm1} that one can constrain the field to propagate strictly in tree diagrams, 
in
\begin{equation}
\int [d \phi \, d \lambda \, dc \, d\bar{c}] e^{i \tilde{S_1}}
\end{equation}
using the modified action
\begin{equation}
\tilde{S_1}(\phi, \lambda, c, \bar{c}, J, \Lambda) = \int d^4 x \left( {\cal L} + \lambda \frac{\delta S_1}{\delta \phi} + 
{\bar c} \frac{ \delta^2 S_1}{\delta \phi^2}c + J\phi + \Lambda \lambda \right)
\end{equation}
that inserts the constraint with a delta function $\delta(\phi - \phi_c)$ in the path integral (where $\phi_c$ is the 
solution of $\delta S /\delta\phi =0$) generalizing and improving previous work by \cite{gozzi}.

The Lagrange multiplier, $\lambda$, enforces the constraint and provides the linear terms that are crucial in the analysis, while the ghosts $c, \bar{c}$ cancel the one loop contributions. Because of the linear terms, and the combinatorics which will be described shortly, there are no higher loop terms, and the propagation is ``classical", in tree diagrams.

Inversion of the kinetic terms yields the propagators
\begin{equation}
\int [d \phi d \lambda] e^{i \int (\frac{1}{2}\phi K \phi + \lambda K \phi +J \phi 
   +\Lambda \lambda)} = N e^{-\frac{i}{2}\int (2 J G \Lambda - \Lambda G \Lambda)}
\end{equation}
where $K=-(\partial^2 + m^2)$ is the kinetic term
and $G=1/(k^2 - m_1^2 + i \epsilon)$ is the Feynman propagator
for a scalar field with mass $m$.
There is no $\phi-\phi$ propagator, but it is reconstructed in the Feynman diagrams with
a mixed $\lambda - \phi$ propagator equal to $G$
and a $\lambda -\lambda $ propagator equal to $-G$.

Each vertex from $\cal{L}$ has an associated vertex from $\lambda \frac{\delta S}{\delta \phi}$ with a $\phi$ leg replaced by $\lambda$, and the ``classical", tree diagrams are reconstructed with these Feynman rules as in Fig.~1. There is an odd number of Feynman diagrams of the form of the first diagram, with alternating signs, that add to the second diagram of the classical theory (which does not exist here, since there is no $\phi-\phi$ propagator).

The important advantage of the formalism described so far is that it allows generalizations to the case where there are interactions with other, purely quantum, unconstrained fields. 

The action $S_2 =\int d^4 x ({\cal L}_1+ {\cal L}_2 + {\cal L}_{\rm int})$, with ${\cal L}_2 =
 \frac{1}{2}(\partial_\mu \psi)^2 -\frac{1}{2} m_2^2 \psi^2 - \frac{g'}{4!}\psi^4$
and ${\cal L}_{\rm int} = -\frac{g''}{2}\phi^2 \psi^2$
can also be modified to
\begin{equation}
\tilde{S}_2 = S_2 + \int \lambda \frac{\delta S_2}{\delta \phi} + \bar{c} \frac{\delta^2 S_2}{\delta \phi^2} c +
J\phi +J'\psi +\Lambda \lambda
\end{equation}
and the Feynman rules derived in the usual manner will constrain the $\phi$ field to propagate classically as above, while also interacting with the purely quantum $\psi$ field, as in the diagram of Fig.~2, where the solid lines that represent the $\phi$ field propagate ``classically", in a tree ``skeleton" while interacting with the wavy lines of the $\psi$ field with the usual quantum behavior. The solid lines ($\phi-\phi$ propagators) as explained before and in Fig.~1, do not exist in the theory,
but are reconstructed with the Feynman rules of the modified theory.

The effective action of the theory can also be obtained with these rules, but there is a difference compared with the usual calculations of quantum field theory. The result given in \cite{dkm1} gives the one-loop contribution for the effective potential  before renormalization as
\begin{equation}
\frac{i}{2}\int \frac{d^4 k}{(2\pi)^4}\ln \left(\frac{(k^2 - \tilde{m}_1^2)(k^2 - \tilde{m}_2^2) + (g''\phi\psi)^2}
{(k^2 -\tilde{m}_1^2)}\right)
\end{equation}
The numerator in this expression represents the usual, one-loop effective potential, when both $\phi$ and $\psi$ are quantum fields, $\tilde{m}_1^2 = m_1^2 +\frac{1}{2} g \phi^2 + g'' \psi^2$, $\tilde{m}_2^2 = m_2^2 +\frac{1}{2} g' \psi^2 + g'' \phi^2$
and the denominator subtracts the one-loop contribution of the $\phi$ field with the ghost ($\bar{c}c$) determinant.

The one-loop contribution cannot be obtained with the usual methods of \cite{jackiw}, \cite{bfm}, which use the quadratic terms and expand around a background, discarding the linear terms, since now the linear terms are important in enforcing the constraint.
The calculations need to be done in the original manner of \cite{coleman} and subtractions are needed.
In higher loops the calculations are done with the relevant, one-particle irreducible diagrams, using the dressed propagators.
The point of this formalism is that the needed subtractions are done consistently.

\section{The cubic interaction in six dimensions}

In this Section, I will apply the formalism to a theory of an $O(N)$-symmetric quantum scalar field, $\psi^a$, $a=1,...,N$, interacting with a ``classical" field, $\phi$, in six spacetime dimensions, with
\begin{equation}
{\cal L} = \frac{1}{2} (\partial_\mu \psi^a)^2 +\frac{1}{2}(\partial_\mu \phi)^2 +\frac{g_1}{2} \phi \,\psi^a\psi^a+
\frac{g_2}{6} \phi^3.
\end{equation}
The fully quantum theory was studied in \cite{klebanov} and IR stable fixed points were found for large values of $N>1038$ that matched with lower-dimensional CFT's.
Radiative symmetry breaking,
dimensional transmutation and the Coleman-Weinberg mechanism with the associated metastable vacuum were also studied in \cite{souza}.

Here, I will be interested in the case where one field is constrained to propagate classically (it is the field that would originate from 
the Hubbard-Stratonovich transformation in the large $N$ approach).
The matching now is not expected to be exact beyond the leading order in the $1/N$ expansion,
but in the renormalization group analysis the stable infrared fixed points appear for much smaller values of $N$ in the $\epsilon$ expansion.
I also consider the Coleman-Weinberg mechanism that appears from the quantum field in six spacetime dimensions, creating a metastable vacuum for the ``classical" field.

The action, $S=\int d^{d} x \, {\cal L}$, is modified in the same manner as before, in order to treat the
$\phi$-field classically, and essentially cancel the scalar loops involving it.

The modified action,
$\tilde{S}= S+\int d^d x (\lambda \frac{\delta S}{\delta \phi} + \bar{c} \frac{\delta^2 S}{\delta \phi^2} c +
J\phi +J'_a\psi^a +\Lambda \lambda) $,
enforces the same behavior as before, 
and cancels loops of the $\phi$-field consistently.
At one-loop, we get the beta functions
\begin{equation}
\beta_1 = -\frac{\epsilon}{2} g_1 + \frac{(N-8) g_1^3 - 12 g_1^2 g_2}{12 (4\pi)^3},
\label{b1}
\end{equation}
\begin{equation}
\beta_2 =-\frac{\epsilon}{2} g_2 +\frac{ N (g_1^2 g_2 -4 g_1^3)}{4 (4\pi)^3},
\label{b2}
\end{equation}
for $g_1$ and $g_2$, respectively, in $d=6-\epsilon$ dimensions,
and
\begin{equation}
\gamma_\phi = \frac{1}{(4\pi)^3}\frac{N g_1^2}{12}, \,\, \,\,\,\,  \gamma_\psi =\frac{1}{(4\pi)^3}\frac{g_1^2}{6}
\label{g12}
\end{equation}
for the anomalous dimensions of $\phi$ and $\psi^a$.

In order to examine the one-loop fixed points in the $\epsilon$ expansion we set
\begin{equation}
g_1= \sqrt{\frac{6 \epsilon (4\pi)^3}{N}}\, x,\,\,\,\,\,\,\,
g_2= \sqrt{\frac{6 \epsilon (4\pi)^3}{N}} \,y,
\end{equation}
and solve the equations for the zeros of the $\beta$ functions,
\begin{equation}
N x = (N-8)x^3 -12 x^2 y,
\end{equation}
\begin{equation}
N y = 3 N x^2 y - 12 N x^3.
\end{equation}

These equations are easily solved analytically as
\begin{equation}
y = \frac{12 x^3}{3 x^2 -1},
\end{equation}
\begin{equation}
x^2 = \frac{ 2(N-2) \pm \sqrt{N^2 +152 N +16}}{3 (N-56)}.
\end{equation}

The origin $(x, y)=(0, 0)$ is a UV stable fixed point for all $N$.
For $N> 56$ there are four real solutions, the two solutions in the first and third quadrant are IR stable fixed points,
and the two solutions at the second and fourth quadrant are saddle points.
For $N\leq 56$ the two stable fixed points disappear (they become purely imaginary solutions) and only the two real saddle points remain.

I show the diagram for the zeros 
and the flows of the $\beta$ functions for $N=80$ in Fig.~3, and for $N=40$ in Fig.~4, and we can compare the results here with \cite{klebanov} and see that the flows are similar and the stable fixed points appear for smaller values of $N$.
The functions $f(x, y)$ and $g(x, y)$ of the plots correspond to $\beta_1$ and $\beta_2$, respectively.
As was mentioned earlier, the matching of the anomalous dimensions of the fields with the lower dimensional CFT, is not exact 
beyond the leading order in the $1/N$ expansion. This is expected since the theory is modified. 
It is believed, however, that the results obtained from this method can provide insights for possible appearance 
of ``classical" behavior and other interesting phenomena in UV completions of quantum field theories \cite{emp1, emp2}.

Radiative symmetry breaking is also expected from power counting for the cubic interaction in six spacetime dimensions, but the emerging vacuum is now metastable, since the energy is not bounded below. It was shown for the fully quantum theory in \cite{souza}, but similar results are expected when only the $\psi$ field gives the loop contributions, while the $\phi$ field is treated classically, as above. The effective potential for $\phi$ can be written as 
\begin{equation}
V_{\rm eff} (\phi) = \frac{1}{6} g_2 \, \phi^3 \, F(g_1, g_2, \ln\frac{\phi^2}{\mu^2}),
\end{equation}
and treated with the renormalization group method at the scale $\mu$ \cite{coleman, ford}
\begin{equation}
(\mu \frac{\partial}{\partial\mu} + \beta_1 \frac{\partial}{\partial g_1}
  +\beta_2 \frac{\partial}{\partial g_2} + \gamma_\phi \frac{\delta}{\delta \phi} ) \,  V_{\rm eff} (\phi) =0.
\end{equation}

After using the relevant expressions from (\ref{b1}, \ref{b2}, \ref{g12}) in six dimensions (with $\epsilon =0$)
we find, in the leading logarithm approximation, 
\begin{equation}
V_{\rm eff} (\phi) = \frac{1}{6} \phi^3 \left(g_2 +
\frac{N(g_1^2 g_2 -2 g_1^3)}{256 \pi^3} (\ln \frac{\phi^2}{\mu^2} - \frac{11}{3})\right)
\end{equation}
with renormalization, as in \cite{coleman}, $V'''_{\rm eff} (\phi) |_{|\phi|=\mu} = g_2$.

Now,  we can examine dimensional transmutation, as in the previous works,
since we have a well-defined (although metastable) vacuum (minimum of the effective potential)
for $\phi$, coming strictly from $\psi$-loops, using the combinatorics described in this work.

After imposing
$V'_{\rm eff} (\phi) |_{\phi =-\mu} = 0$, we get 
$
g_2 = 3 N c (g_1^2 g_2 -2 g_1^3),
$
with the constant $c=1/(256 \pi^3)$, and we can solve for
$
g_2 = \frac{- 2 N c g_1^3}{1 - N c g_1^2},
$
in the usual spirit of dimensional transmutation, 
``trading parameters", from the initial couplings $g_1$ and $g_2$,
to $g_1$
and the generated mass of the $\phi$ field at the metastable minimum,
$
V''_{\rm eff}(\phi) |_{\phi =-\mu} = 2 c N g_1^3 \mu.
$

The quantum field $\psi^a$  was considered here in the symmetric phase, although generalizations are possible, and the results found agree with \cite{souza}, with the modifications regarding the classical field taken into account.

\section{Comparison with other works}

The work presented here, originally in \cite{dkm1}, is a generalization, and also an improvement,
of work by the groups of  \cite{gozzi}, which consider the action
\begin{equation}
\bar{S} = \lambda \frac{\delta S}{\delta \phi} +\bar{c} \frac{\delta^2 S}{\delta \phi^2} c,
\label{gozzi}
\end{equation}
as opposed to the action
\begin{equation}
\tilde{S} = S + \lambda \frac{\delta S}{\delta \phi} +\bar{c} \frac{\delta^2 S}{\delta \phi^2} c,
\label{metaxas}
\end{equation}
proposed in \cite{dkm1}.

It is easy to derive the Feynman rules from (\ref{gozzi}), which have the mixed, $\lambda-\phi$ propagators,
but not the $\lambda-\lambda$ propagator, which gives a negative contribution, and 
leads to the classical diagrams described in Fig.~1. 
These simple classical scattering diagrams, and many other diagrams, described here, cannot be obtained with the methods of \cite{gozzi}.
There were some attempts to reconcile their work with what can be considered ``classical" \cite{gozzi2},
modifying the derivation of the Feynman rules,
but the usual way to derive propagators from the path integral \cite{zee} does not agree.

What is more important, the formalism of \cite{dkm1}, based on (\ref{metaxas}),
allows the generalization to the case where there are interactions of the ``classical" field
(which is, in any case, something artificial, that needs to be justified)
with the usual quantum fields that exist in Nature.

The work of \cite{gozzi}, based on (\ref{gozzi}), 
has some interesting insights on classical mechanics, especially with respect to symmetries,
but it does not generalize to the quantum treatment.

There are other works that investigate interactions between fields with classical behavior and quantum fields
\cite{son}. Usually, these works examine separations of scales, which are relevant and useful for many physical situations.
Generally, however, and formally, separations of scale are physically not exact, and adjustments need to be done ``by hand".
The work of \cite{dkm1}, presented in more detail here, gives a formal and consistent treatment, in terms of Feynman rules,
that can easily and consistently be generalised to more fields, quantum or classical, higher loops, etc.

On the other hand, purely classical fields do not exist in Nature, so the use of the techniques described here needs to be 
justified by physical reasons in order to be applied to relevant problems.

\section{Comments and applications}

The results, comments, and applications of the formalism presented here and in \cite{dkm1} are somewhat disparate, since
the main result is technical and combinatorial, so it can be applied to various problems.
Its main importance lies in the treatment of linear terms in the perturbation expansion. These are usually discarded,
for good and justified reasons,
since most problems involve perturbations around a stable minimum of a functional, and use 
harmonic oscillators and their generalizations in order to describe these fluctuations.
In these cases, the usual methods of \cite{jackiw}, \cite{bfm}, are consistently applied, and possible ``tadpole" terms,
which may arise by shifting the fields, can be consistently set to zero.

The linear terms considered here are related to different effects, and cannot be set to zero in a consistent manner. 
An appearance of such linear terms is in 
the Lagrange multipliers used to describe a constraint, and the first application was to the treatment of 
an interaction as ``classical", propagating in tree diagrams.

It should be noted that, in order to reconstruct the full classical theory, and settle problems
of consistency and unitarity, the full set of tree diagrams should be considered.
Perhaps a more useful and relevant investigation would be to the possibility of 
emergence of the gravitational or another interaction as an effective classical, or partly classical, theory.

Another application, which was pursued in subsequent works \cite{dkm2, dkm3},
involved the treatment of the constraints of the non-Abelian gauge theory, and led to 
the description of the vacuum of the strong interaction, with numerous applications.

A conceptual problem, which was also a motivation of the initial work,
is the measurement problem in the various interpretations of quantum mechanics.
It is sometimes said that the measurement process involves the interaction
of a quantum system (to be measured) with a classical system (the detector), and it was hoped that the formalism presented here
would be useful towards this description. In fact, the simple examples presented here and in \cite{dkm1}
can be seen in this way. A classical field (representing a measuring device) interacts with a purely quantum field,
and develops a vacuum expectation value (corresponding to a measurement).
This is very rudimentary, and has not yet been formulated in a more practical example.
Also, results from other investigations reveal another possible connection
with the measurement problem, to be discussed in a related work \cite{dkm4}.

Finally, as far as other possible applications are concerned, it should be mentioned that it is possible to extend the formalism to fermion and higher spin fields, and also investigate non-renormalizable interactions and the structure of the divergencies when the perturbation theory is re-organized in the manner described.

\section*{\centering Note}
 There are no additional data and no conflicts of interest regarding this work. The Feynman diagrams were made with Jaxodraw \cite{jaxo}, and the graphs with the free version of Google Gemini.

\newpage

\begin{figure}
\centering
\includegraphics[width=60mm]{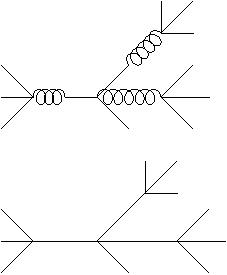}

\caption{   The top diagram is one of the many with the same line structure that can be formed with the Feynman rules described in the text.
An odd number, with alternating signs, can be formed, and their sum gives the bottom tree diagram of the ``classical" theory 
(solid lines denote $\phi$ and wavy lines denote $\lambda$).}
\end{figure}

\begin{figure}
\centering
\includegraphics[width=60mm]{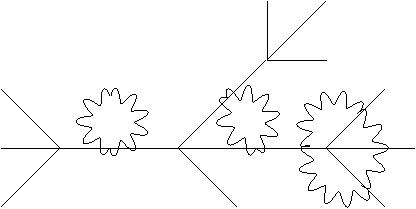}

\caption{A diagram of the interacting quantum-classical theory, with the classical field propagating in tree diagrams, with arbitrary loops of the quantum field (wavy lines). The classical tree skeleton is formed with summations of mixed propagators as in Fig.~1. }
\end{figure}

\newpage

\begin{figure}
\centering
\includegraphics[width=150mm]{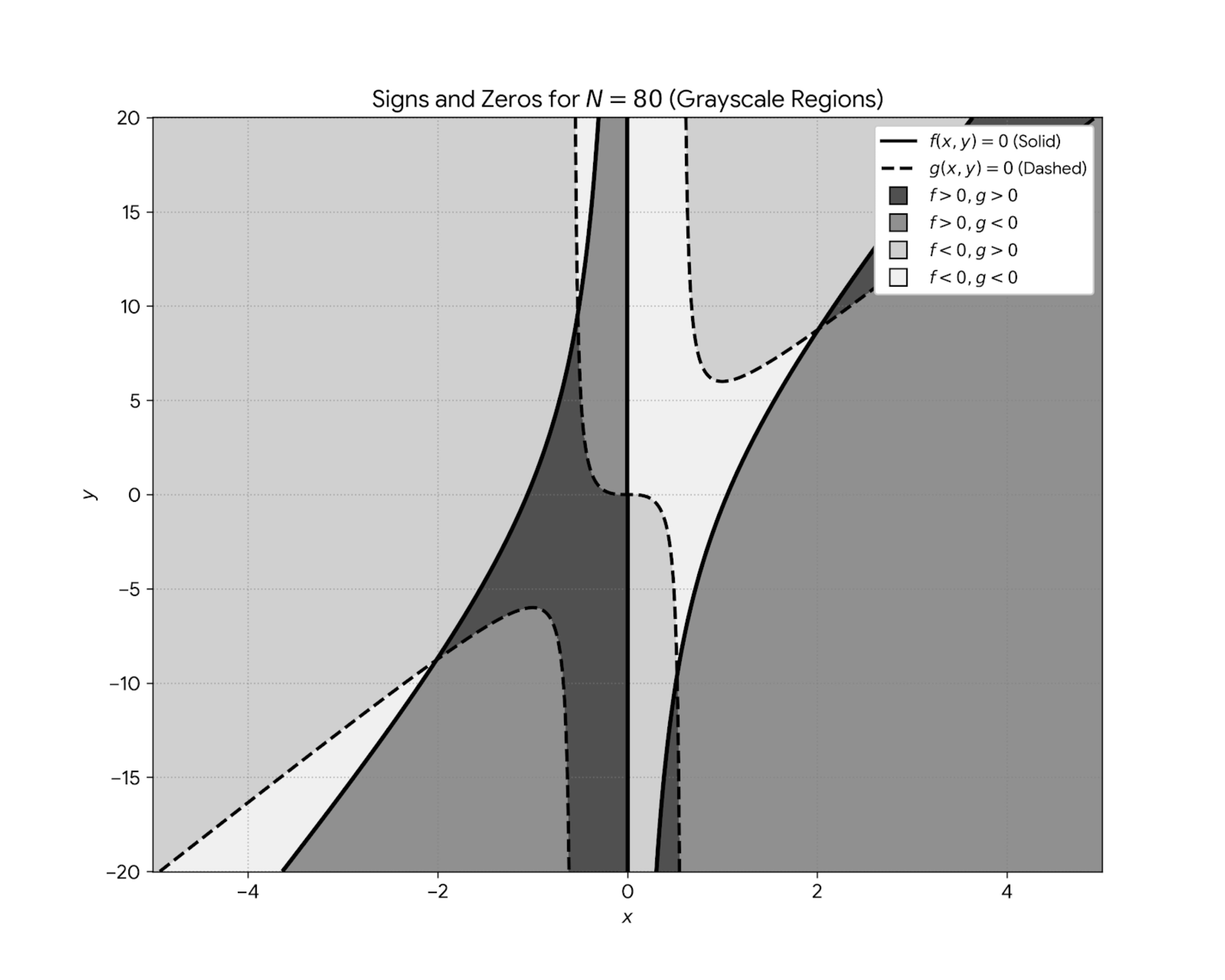}

\caption{ The signs and zeros for the beta functions, for $N=80$. }
\end{figure}

\begin{figure}
\centering
\includegraphics[width=150mm]{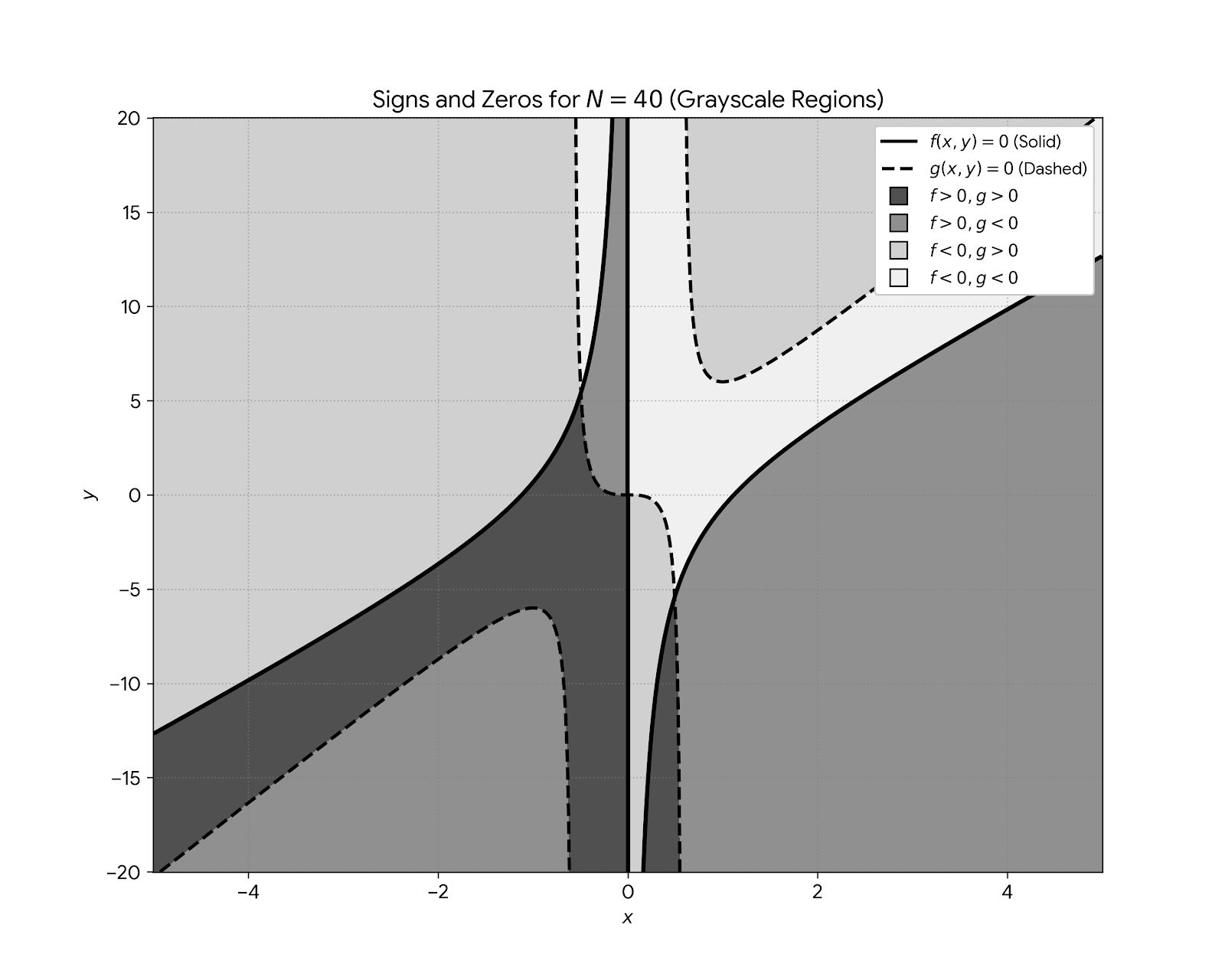}

\caption{  The signs and zeros of the beta functions for $N=40$. }
\end{figure}

\end{document}